# Some Useful Properties of Rotational Gibbs States In Chiral Conformal QFT

Bert Schroer
Institut für Theoretische Physik
Freie Universität Berlin
Arnimallee 14,   D-14195 Berlin

April 1994

## Abstract

We derive the Kac Wakimoto formula in the algebraic framework of chiral conformal QFT s a consequence of the universal behaviour of high-temperature Gibbs-states on intertwiner subalgebras. By studying selfintertwiners in the presence of other charges, we obtain charge-polarization coefficients which generalize the entries of Verlinde matrix S and belong to higher genus mapping class group matrices.

# 1 Introduction

Although the basic physical assumption in the algebraic approach to quantum field theory are formulated in terms of nets of local von Neumann algebras [1], it has been convenient and helpful to introduce specific global observable algebras.

The Doplicher-Haag-Roberts approach uses a net indexed by finite Minkowski-space double cones and directed towards spatial infinity. The ensuing inductive $C^*$ limit was called the quasilocal algebra $\mathcal{A}$ [1]. Among its many physically useful concepts was the construction of (outer) charge creating endomorphism and the ensuing superselection sectors as a spatial limit of (inner) charge transport [1] (Splitting charges and dumping the anti-charge "behind the moon".)

A quite different global algebra arises if there are physical requirements which enforce a space-time compactification (viz. in conformal QFT or 3-d theories with semiinfinite string-like charges). In that case the globalization has to be done by a universal construction [2]. The new $C^*$ algebra $\mathcal{A}_{uni}$ [3] is in the presence of nontrivial superselection sectors genuinly bigger than $\mathcal{A}$ (but fortunately not much bigger).

$$\mathcal{A} \subset \mathcal{A}_{uni} \qquad (1)$$

In chiral conformal QFT the latter results by removing the point $\infty$ from $\mathcal{A}_{uni}$. This removal does not only disturb the net (the intervalls containing $\infty$ are removed), but also makes the global algebra smaller (one looses the global selfintertwiners as well as the center of $\mathcal{A}_{uni}$). This inclusion is maintained for the von Neumann algebras obtained through particular represenation, e.g.

$$\pi_0(\mathcal{A}) \subseteq \pi_0(\mathcal{A}_{uni}). \qquad (2)$$

the equality in (2) being valid iff the original net was "strongly additive", [4].

The global selfintertwiners which account for the bigger size of the new universal observable algebra give rise to abelian charges $C_\rho$ whose values in the various irreducible representation $\pi_0 \cdot \sigma$ are known to define the entries into Verlindes Matrix $S$ [3].

This raises the question whether the universal algebra could lead to a different way of keeping track of intertwiner algebras, Markov traces on the braid group algebra $CB_\infty$ and perhaps also to generalizations of the Verlinde matrix $S$ (going into the direction of mapping class group matrices).

The formula of Kac-Wakimoto [3], which relates the quantum dimension of loop group representations to the asymptotic behaviour for large temperature partition function illustrates what we are aiming at.

In this paper we will derive the Kac-Wakimoto formula solely by using the physical principles and the von Neumann algebraic methods of algebraic QFT. It turns out that, by using a high-temperature universality principle, all the information using left inverses in the DHR approach can be replaces by studying infinite temperature limits of rigid-rotation Gibbs states. These states, already at finite temperature, do not depend on the particular localization of the endomorphism, but only on their equivalence class i.e. on the sector.

The question of "higher genus" generalization of $S$ is positively answered by studying the global selfintertwining of one charge in the presence of many spectator charges (and not just its anticharge). The resulting multicharge polarization coefficients are indeed entries of higher genus mapping class group matrices, but the physical interpretation remains somewhat mysterious.

## 2   Global Charges and Asymptotic Properties of Gibbs States in Chiral Confromal QFT

In the previous paper [3] we constructed invariant charge operators ("charge measures") $C_\rho$ by first splitting the vacuums into charge-anticharge endomorphisms $\rho\bar\rho$ with the help of an intertwiner $R\epsilon\mathcal{A}_{uni}$:

$$R_\rho id(A) = \rho\bar\rho(A)R_{\bar\rho} \qquad \forall A\epsilon\mathcal{A}_{uni} \qquad (3)$$

then transporting $\rho$ "once around":

$$V_\rho \rho(A) = \rho(A)V_\rho, \qquad (4)$$

$V_\rho = 2\pi$-charge transport (expressible in term of a rotation cocycle [3]
and finally recombining the $\rho\bar\rho$ representations to the identical (vacuum) representation. Putting the three steps together we obtain the invariant charge operators $C_\rho$

$$\begin{aligned} C_\rho &= R^*_{\bar\rho} V_\rho R_{\bar\rho} = C_{[\rho]} \\ [\rho] &= \text{superselection equivalence class of} \quad \rho. \end{aligned} \quad (5)$$

These matrices are well-known and have been obtained by evoking a picture of conformal QFT's on Riemann surfaces [10], but they also appear in pure combinatorial approaches to invariants of 3-manifolds. We emphasize the "polarization coefficient" picture, because we consider this quantum physical understanding as an important ingredient of the still elusive "quantum symmetry". It is this new quantum symmetry which should explain the usefulness of Riemann surfaces and 3-manifolds for low dimensional quantum physics and not the other way around [13]. At the time of the Moore Seiberg work a good understanding of $\mathcal{A}_{univ.}$ and central charges was not yet available.

Making the non-degeneracy assumption that the charge measures can be distinguished by their charge values in the various charge sectors, one arrives at Verlindes matrix $S$ as the collection of charge values:

$$S_{\rho\sigma} 1 = \frac{d_\sigma}{\sqrt{\sum_i d^2_{\rho_i}}} \pi_0 \sigma(C_{\bar\rho}) \quad , \quad d_\sigma = \text{statistical dimension} \quad (6)$$

Here we use the standard notation $\bar\rho$ for the conjugate endomorphism of $\rho$. For our purpose the most relevant objects are the central projectors:

$$P_\rho = \sum_\rho \frac{S_{\rho\sigma}}{S_{0\sigma}} C_\sigma \quad , \quad P_{\rho_i} P_{\rho_k} = \delta_{\rho_i,\rho_k} P_{\rho_i}. \quad (7)$$

They decompose the universal algebra according to

$$\mathcal{A}_{uni} = \sum_i P_{\rho_i} \mathcal{A}^{(\rho_i)}_{uni}. \quad (8)$$

In the DHR theory, which applies to $4d$ QFT, one also finds such a central decomposition if one uses a suitably compactified observable algebra (by extending the net through "cones at infinity" [12]). But in that case there are no selfintertwining charge transporters leading to central charges $C_\rho$, i.e. the central projectors have no geometrical significance. The presence of such DHR central projectors in our $\mathcal{A}_{uni}$ has been excluded by our non-degeneracy assumption [3].

The main formula of the DHR theory is however the decomposition of the vacuum representation of the DHR field algebra $\hat{\pi}_0$ restricted to the observable algebra:

$$\hat{\pi}|_{\mathcal{A}} = \sum_{\xi} 1_{\xi} \otimes \pi_{\xi}(\mathcal{A}) \quad (9)$$

where $1_{\xi}$ are the identities in the irreducible representation spaces of a compact group. Clearly each irreducible representation of the observable algebra occurs precisely with the representation dimension $d_{\xi}$ and hence the decomposition (9) is often written in the multiplicity form:

$$\hat{\pi}|_{\mathcal{A}} = \sum_{\xi} d_{\xi} \pi_{\xi}(\mathcal{A}) \quad (10)$$

An analogous multiplicity theory with noninteger statistical dimension for our circular chiral observable algebras $\mathcal{A}_{uni}$ (and more generally for low dimensional QFT) is presently not known. It turns out however, that the central projectors in our case can be viewed as projectors of a type $II_1$ von Neumann factor which have a unique natural tracial state and furnish a (noninteger) dimension [see section 3]

$$tr P_i = \frac{d_i}{d} \quad , d = \sum d_i \quad (11)$$

This analogy to the standard case of a group symmetry gives confidence that despite the present lack of a field algebra (and therefore of the origin of formula (4)), the global structure of the universal algebra yields an important step towards an understanding of quantum symmetry. In this note we will add two recent results which strengthens this view:

1) A model-independent derivation of the Kac-Wakimoto [6] formula for statistical dimensions in terms of large temperature limits:

$$\lim_{\beta \to 0} \frac{tr e^{-\beta L_0^{(\rho)}}}{tr e^{-\beta L_0^{(\sigma)}}} = \frac{d_{\rho}}{d_{\sigma}} \quad (12)$$

2) A study of the algebra generated by matrix elements of global selfintertwiners in the "path-basis". This algebra leads to invariant "polarization charges" which are proportional to the already mentioned central charges $C_\rho$ with the polarization coefficients describing mapping class matrices for higher genus.

Our approach to the second problem is opposite to the one usually taken in the literature, but complies with reference [3]. Instead of using field theoretic ideas to construct invariants of 3-manifolds and mapping class group representations, we stay with Minkowski-space QFT and consider our geometrical findings as a surprising aspect of a new quantum symmetry concept which puts severe limits on a separation of space-time symmetries from the "would be" internal symmetries and hence makes low-dimensiona QFT significantly different from higher dimensional QFT, for which the two kind of symmetries (apart from the $2\pi$-rotation for spinors) form a tensor product [1].

## 3   Kac-Wakimoto Formula and Relative Size from first Principles

Consider a universal algebra $\mathcal{A}_{uni}$ on $S^1$ with a finite number of nondegenerate superselection sectors $[\rho_i] i = 1 \ldots n$. In that case the affiliated central charges $C_{\rho_i}$ span the center of $\mathcal{A}_{uni}$. Define a localized "master endomorphism" [8]

$$\hat{\rho}(A) := \sum_i W_i \rho_i(A) W_i^* \quad , \quad A \in \mathcal{A}_{univ.} \tag{13}$$

by choosing the localization of all $\rho_i$ inside one proper interval $\rho_i \in (I)$, and intertwining orthogonal isometries $W_i$ localized in a slightly bigger proper interval of $S^1$. It is known that such a master endomorphism can be chosen as a selfconjugate sector: $\hat{\rho} = \bar{\hat{\rho}}$. [8]

Let $\hat{\phi}$ be the unique standard left inverse of $\hat{\rho}$, [3] i.e. a positive map of $\mathcal{A}_{uni}$ with

$$\begin{aligned} \hat{\phi} \cdot \hat{\rho} &= id \\ \hat{\rho} \cdot \hat{\phi} &= \varepsilon_{min} \quad , \\ \varepsilon_{min} &= \text{minimal conditional expectation of } \mathcal{A}_{uni} \to \hat{\rho}(\mathcal{A}_{uni}) \end{aligned} \tag{14}$$

This unique left inverse can be represented in the well-known form (using the self conjugacy of $[\hat{\rho}]$ and choosing $\hat{\rho} = \bar{\hat{\rho}}$:

$$\hat{\phi}(A) = R^*\hat{\rho}(A)R \ , \quad \forall A \epsilon \mathcal{A}_{uni} \tag{15}$$

with $R$ a suitable intertwiner $R \in (\rho^2, id)$. The iteration of this left inverse evaluated in the vacuum representation $\pi_0$ gives a tracial state on the centralizer algebra of $\hat{\rho}$:

$$tr := \lim_{m \to \infty} \pi_0 \cdot \hat{\phi}^m \quad \text{is a trace on} \quad M = \cup_m (\hat{\rho}^m, \hat{\rho}^m) \tag{16}$$

The $C^*$ algebra $M$ is known to have a unique trace and its weak closure with respect to this tracial state is the hyperfinite $II_1$ factor. Its restriction to the colored ribbon-braid-groupoid algebra $RB_\infty \subset M$ yields a ribbon-Markov trace [3].

We now define a faithful Gibbs state on $\mathcal{A}_{uni}$ as

$$\omega_\beta(A) = \frac{1}{Z} Tr \pi_0 \hat{\rho}(Ae^{-\beta L_0}) = \frac{1}{Z} \sum_i Tr \pi_0 \rho_i(A) e^{-\beta L_0^{\rho_i}} \tag{17}$$

with $Z = \sum_i Tr \pi_0 e^{-\beta L_0^{\rho_i}}$. Here $\rho_i(e^{-\beta L_0})$ is a short-hand notation for $e^{-\beta L_0^{\rho_i}}$, where $L_0^{\rho_i}$ implements the (infinitesimal) rigid rotation in the representation $\pi_0 \cdot \rho_i$. Since it is believed that $L_0$ is affiliated via an energy-momentum tensor with the $C^*$ algebra $\mathcal{A}_{uni}$, this way of writing may actually be more than a short-hand notation.

The Gibbs state becomes formally a tracial state in the limit of infinite temperature.[1] This limit does not exist on $\mathcal{A}_{uni}$. To be more precise the sequence of states for increasing temperatures pocesses subsequences with limit points (since the state space on a $C^*$-algebra is compact). But the limit states are generally (and in particular in our case) nonnormal (singular) on the net. A good example is the infinite temperature state of the circular Weyl algebra (by abuse of notation) [7]. The representation of the spacetime transformation is lost in this limit. We will now argue that it retaines its good properties if one restricts to the subalgebra $M$ of intertwiners. In fact on each finite dimensional subalgebra $M_m$ the limit exists as a normal state since there are no nonnormal states on finite dimensional algebras. However the proof that it is a regular tracial state requires the following assumption.

---

[1]The unnormalized positive Gibbs functional becomes a "tracial weight" on the type I von Neumann algebra $\pi_{\omega_\beta}(\mathcal{A}_{uni})$ but this weight is not useful if one deals with local properties.

Technical universality assumption:

The infinite temperature limit is (apart from normalizations) independent of the sector. In particular the nucearity-indices [1] of the nuclear maps [1] affiliated with the Gibbs states and the algebra $\mathcal{A}_{univ.}$ are universal.

A strong version of this assumption would be the requirement that the normalized tracial states belonging to different sectors approach each other in the trace norm:
$$\left\| \frac{e^{-L_\rho \beta}}{Z_\rho} - \frac{e^{-L_0 \beta}}{Z_\sigma} \right\|_1 \xrightarrow[\beta \to o]{} 0 \tag{18}$$
Here we only need the weaker version that the partial states, i.e. the Gibbs states restricted to the finite dimension factor algebras $(\hat{\rho}^n, \hat{\rho}^n)$ coalesce in the limti $\beta \to 0$. In the standard setting of chiral conformal field theory, bases on the existence of an energy momentum density affiliated to $\mathcal{A}_{univ.}$, the universality assumption follows from the asymptotic equality of energy level densities belonging to different superselection sectors. We now use the fact that these intertwiner algebras become tracial with the hamiltonian $L_{\hat{\rho}^n}$ (this hamiltonian commutes with these algebras). The weak universality principle then establishes the tracial nature of the Gibbs state $\frac{e^{-\beta L}}{Z}$ in the limit $\beta \to 0$ on all intertwiner algebras and hence on $M$.

As a result of the uniqueness the infinite temperature limit must be proportional (with a possibly diverging normalization factor) to the tracial state $tr$ on $M$ in formula (16) (uniqueness on a factor!)
$$\varphi := \lim_{\beta \to 0} \omega_\beta |_M \quad \propto \quad tr \tag{19}$$

In other words the tracial state of algebraic QFT may be substituted by a more physical limiting Gibbs state. In particular, we obtain for the central projection
$$\lim \frac{\omega_\beta(P_\sigma)}{\omega_\beta(P_\kappa)} = \frac{tr(P_\sigma)}{tr(P_\kappa)} = \frac{d_\sigma}{d_\kappa} \tag{20}$$
where the last equation results from (see 10)
$$\pi_0 \phi(P_\sigma) = \sum_i R^* W_i \pi_0 \rho_i(P_\sigma) W_i^* R \tag{21}$$
$$= R^* W_\sigma W_\sigma^* R = \frac{d_\sigma}{\Sigma d_{\rho_i}}. \tag{22}$$

This asymptotic formula (20) is the Kac-Wakimoto formula.

We conjecture that $M''$ is, up to equivalence, the maximal algebra on which the infinite temperature state is regular. A brief comment on the universality assumption is in order. The assumption is fulfilled in all known chiral conformal models. The nuclearity index $\nu_\beta$ is bounded by the first order exponential

$$\nu_\beta < \alpha \exp(\frac{\beta_0}{\beta}) \qquad (23)$$

and the best constant $\beta_0$ is given in terms of the Virasoro energy parameter (a kind of Casimir effect). The best coefficient $\alpha$ is related (as was shown before) to the normalized trace on the central projectors of $\mathcal{A}_{uni}$. We used the word technical in order to indicate that a future better understanding may allow to weaken this assumption or remove it altogether.

The previous infinite temperature discussion clearly transcends the derivation of the Kac-Wakimoto formula. The latter amounts to the temperature-representation of the central projectors of $\mathcal{A}_{uni}$. But together with the endomorphism $\hat{\rho}$ these central projectors generate infinite algebras of intertwiners which constitute what is often referred to as "topological field theory" since $M$ contains in particular the braid group algebra and permits via the Markov trace to construct coloured ribbon invariants and Kirby move invariants of 3-manifolds [3]. We learn from our consideration that infinite temperature limit states on $M$ destroy the representation theory of space-time symmetry groups and in this way may generate $II_1$ algebras (a very unusual state of affairs in quantum theory). An analogue mechanism, but this time with the singular nature of the states coming from gauge invariance (intuitively: summing over infinitly many gauge copies) is expected to result from topological Chern Simons theories in the following way: first one canonically quantizes the Chern-Simons Lagrangian (e.g. in the time-like gauge and then one converts its formal content into a Weyl-like algebra of one-forms. As in conformal QFT one then chooses a maximal local extension (generated by one-forms of finite angular support) and finally one constructs the universal algebra. The gauge invariant state on that algebra should then have the same effect as the high-T state on the associated conformal algebra (i.e. that with the same superselection structure) namely produce the tracial state on a $II_1$ intertwiner algebra which is the characteristic feature of topological quantum field theory. The simplest nontrivial illustration for such a construction would be the abelian $Z_N$ Chern-Simons theory. Unfortunately only the first step in this interesting program has been carried out [9].

## 4 Subalgebras Generated by Selfintertwiners and Polarization Coefficients

In Section 2 we introduced global selfintertwiners $V_\rho \in (\rho, \rho)$ which are nontrivial in $\mathcal{A}_{uni}$ but become trivial (for irreducible $\rho$) in the vacuum representation $\pi_0$. Their physical interpretation in terms of a counter-clockwise charge transportation around $S^1$ is obvious from their construction. By sandwiching them between the "vacuum-split" intertwiners $R \in (\rho\bar\rho, id)$ (the affiliated projectors $RR^*$ are the Jones projectors of the Jones inclusion defined by $\rho$) one obtains the invariant central charge $C_\rho$.

This construction may be generalized in the following way. We may split the $\rho$ with the help of T-intertwiners and attach the selfintertwiner $V_\sigma$ to a secondary $\sigma$-line (using the standard pictorial representation for endomorphisms and intertwiners [9]).

In this way one obtains a "polarization selfintertwiner" $V_\sigma^{(\rho)} \in (\rho, \rho)$ which depends in addition on the detailed charge transport. Since we took the T-intertwiners and their hermitian adjoints $T^*$ from the basis of intertwiners which leads to a "string-basis" of the centralizer algebras, the splitting and recombination would give Kronecker $\delta$'s if we replace $V_\sigma$ by the identity. In other words we obtain algebra-valued matrix elements of $V_\sigma$. With the help of $R \in (\rho\bar\rho, id)$ one may again convert $V_\sigma^{(\rho)}$ into a central element say $C_\sigma^{(\rho)}$. Its graphical representation (from which one easily reads off the algebraic formula) is given in Appendix Fig. 3.

From this representation one easily concludes that this invariant polarization charges is proportional to the charge $C_\sigma$ ($e_i$ stands for the various source-charge-range edges).

$$C_\sigma^{(\rho)} = S(\rho, \sigma, e_i \ldots) C_\sigma \qquad (24)$$

where the coefficient $S$ is presented by the same picture with the $\sigma$-dot deleted (evaluation of selfintertwiners in the vacuum yields the identitiy operator).

In order to have an intertwiner-calculus which treats all irreducible superselection sectors on the same footing, one should transcribe the formalism of this section with the help of additional intertwiners $W_i$ into the formalism based on the "master endomorphism" $\hat{\rho}$ of the previous section. The dimension of the algebraic path-space for an n-fold charge split if (in terms of the fusion matrices $N^a$)

$$\dim(\hat{\rho}^n, \hat{\rho}^n) = [tr(\vec{N}^2)^{n-1}]^2, \vec{N} = \sum_a N^a N^a \qquad (25)$$

i.e. one obtains Verlinde formula with $n$ =genus.

The generalized selfintertwiners can be used to define a subalgebra $G^{glob.}$ of the centralizer algebra $M = \cup_n(\hat{\rho}^n, \hat{\rho}^n)$ which by definition is generated by the algebraic matrix elements of selfintertwiners i.e. by $\hat{V}_\sigma^{(\rho)}$- polarization selfintertwines. It turns out that the polarization coefficients can be interpreted as entries of higher genus mapping class group matrices thus generalizing the relation between the values of invariant charges in charged states and the $g = 1$ Verlinde matrix $S$. We remind the reader that the higher genus mapping class group can be defined via so-called A-B Dehn-twists. The S-operation if formed from adjacent Dehn-twists on intersecting cycles [10]. We refrain here from a more detailed explanation [2] of how the collection of polarization coefficients are converted into entries of mapping class group matrices since we lack a conceptional understanding of why the polarization coefficients contained in the subalgebra $\mathcal{G}^{glob.}$ should be related to entries of mapping class-group matrices. The appearance of such free discrete groups (in form of unitary representations in "charge space") we consider as an important hall mark of the still elusive "quantum symmetry". In contrast to the interpretation of e.g. Moore and Seiberg, we do not primarily associate this finding to "conformal QFT on Riemann surfaces" but rather consider it as an enrichment of the QFT symmetry problem.

In our interpretation the higher genus aspects of chiral conformal QFT should be viewed as an "operation" which assigns a collection of numbers (the entries of mapping class group matrices) to Riemann surfaces and not as a field theory which "lives on" a particular Riemann surface in the sense of localizability. In other words we expect that the still elusive new quantum symmetry explains why charges and their composition laws have such a natural relation with Riemann-surfaces and 3-manifolds. The observations in this paper belong to "quantum invariant theory". A deeper understanding would involve questions of why nets on a circle not only "feels" the Moebiusgroup as the symmetry group of the vacuum, but also all those discrete Fuchsian subgroups as well as their deformation theory (which can be discribed in terms of mapping class groups). We expect that a field algebra which encodes all these properties must be radically different from the standard one [13]. On the physical side the new symmetry and its field-carriers should trivialize the classification problem of e.g. chiral conformal QFT and that of "free" anyons and plektons. In this way one could hope to bridge the present gap between the rich theoretical numbers (statistics- or spin phases, statistical dimensions) and their potential experimental counterparts (rational numbers of the electromagnetic fractional Hall-effect, amplification-factors in thermodynamical $T_c$-gap relations, which measure the size of degrees of freedom carried by new quasi-particles).

---

[2] The interested reader is referred to the pictures in [11] which agree with ours. Their meridian constructions correspond to our global selfintertwiners.

# 5 Outlook

The high temperature behaviour of Gibbs state and its relation to tracial states on intertwiner algebras (topological quantum field theory) and the Kac-Wakimoto formula was more or less to be expected. Nevertheless the encoding of fundamental inclusion invariants (statistical dimensions, Markov traces) in rather simple physical Gibbs-states (instead of left inverses and conditional expectations) is a quite useful enrichment of field theoretic understanding.

The appearance of higher genus data as polarization coefficients of generalized central charges is however somewhat surprising. It is a consequence of the existence of central charges in the universal algebra. Wheras in the DHR theory one can "create" charges by transport from infinity leaving the anti-charge behind at infinity (this is the physical picture for the charge creating endomorphism [1]), the universal algebra only allows for charge transport "once around" leading to the central charges $C_\rho$. A field-theoretic understanding of "duality" requires to combine these two algebraically mutually exclusive ideas. Fortunately there is a third form of charge transport related to the Connes cocycle and their use for the creation of charged states on subalgebras. Using the so-called natural cone description of states, this construction has an interpretation in terms of "dumping" the anticharge into the commutant. We hope to show in a future paper how "duality" arises from an interplay of the temperature theory (described in this paper) and a new euclidean theory which has a different ∗-operation and which allows for a charge transport which leads to charge creation.

I am indebted to M. Karowski and H.W. Wiesbrock for fruitful discussions.

# 7  Figure Captions

Fig. 1: Graphical representation of global selfintertwiners and its "flip"- property.

Fig. 2: Polarization intertwiners $V_\sigma^{(\rho)}$. The dependence on the charge splitting edges $e_i$ has been surpressed in order to keep the operator notation simple.

Fig. 3: Invariant polarization coefficients obtained by short circuiting the polarization intertwiners of Fig. 2 with the help of the vacuum-$\rho\bar\rho$ split intertwiners $R$. These coefficients form the building blocks for the higher genus matrices S.

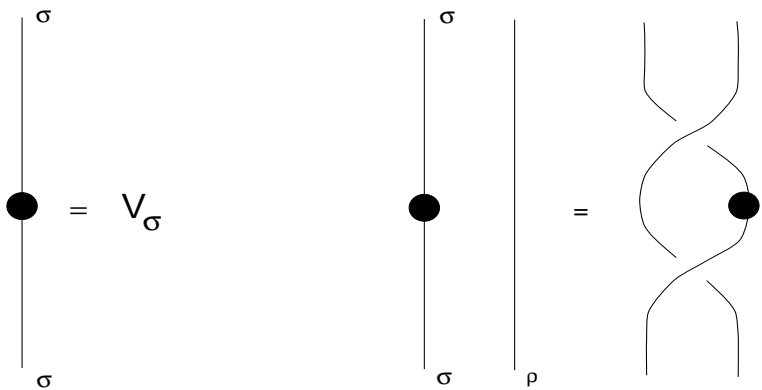
Fig. 1

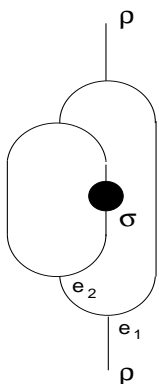
Fig. 2

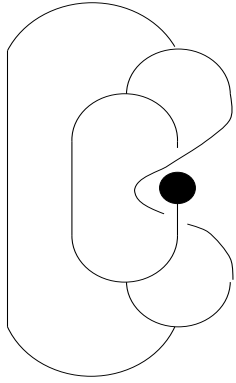
Fig. 3